\documentclass[10pt,twocolumn,twoside]{article}

\usepackage[utf8]{inputenc}
\usepackage[T1]{fontenc}
\usepackage[english]{babel} 
\usepackage{CJKutf8} 

\usepackage{graphicx}
\usepackage{amsmath,amssymb,amsfonts}
\usepackage{float}
\usepackage{xcolor}
\usepackage{titlesec}
\usepackage{authblk} 
\usepackage{fancyhdr}
\usepackage{abstract}
\usepackage{tipa} 
\usepackage[numbers,sort&compress]{natbib}
\usepackage{orcidlink}
\usepackage{hyperref}

\definecolor{pnasblue}{RGB}{0, 114, 188}
\hypersetup{colorlinks=true, linkcolor=pnasblue, citecolor=pnasblue, urlcolor=pnasblue}

\usepackage[left=1.5cm,right=1.5cm,top=2cm,bottom=2cm]{geometry}

\titleformat{\section}{\large\bfseries\color{pnasblue}\sffamily}{\thesection}{1em}{}
\titleformat{\subsection}{\normalsize\bfseries\sffamily}{\thesubsection}{1em}{}

\title{\huge \bfseries \sffamily Implementing Substance Over Form: A Novel Metric for Taxing E-commerce to Address Deterritorialization  \\ \Large \color{gray} \huge \textnormal{践行实质课税原则：一种针对电商脱域化的新型计税指标}}

\author[a,b,1]{Li Tuobang 李拓邦 \orcidlink{0000-0002-2257-2603}}
\affil[a]{Independent Researcher, Zhaoqing, China 独立研究员，肇庆，中国}
\affil[b]{University of California, Berkeley, USA 加州大学伯克利分校，美国}

\affil[1]{To whom correspondence should be addressed. 联系方式。E-mail: lituobang@hotmail.com}

\newcommand{\acknow}[1]{\section*{Acknowledgment 致谢} {\small #1}}

\begin{document}

\begin{CJK*}{UTF8}{gbsn}

\twocolumn[
  \begin{@twocolumnfalse}
    \maketitle
    \begin{abstract}
Against the backdrop of e-commerce restructuring consumption patterns, last-mile delivery stations have substantially fulfilled the function of community retail distribution. However, the current tax system only levies a low labor service tax on delivery fees, resulting in a tax contribution from the massive circulating goods value that is significantly lower than that of retail supermarkets of equivalent scale. This disparity not only triggers local tax base erosion but also fosters unfair competition. Based on the "substance over form" principle, this paper proposes a tax rate calculation method using "delivery fee plus insurance premium" as the base, corrected through "goods value conversion." This method aims to align the substantive tax burden of e-commerce with that of community retail at the terminal stage, effectively internalizing the high negative externalities of delivery stations through fiscal instruments, addressing E-commerce Deterritorialization.
    
在电子商务重构消费模式的背景下，末端快递站已实质上履行了社区零售分销的功能。然而，现行税制仅对其派送费征收低额劳务税，导致其流转的巨大货值在末端环节产生的税收贡献远低于同等规模的零售超市。这不仅引发了地方税基流失，更造成了不公平竞争。本文基于实质课税原则，提出一种以“派费+保费”为基数、通过“货值换算”进行修正的税率计算方法，旨在实现电子商务业与社区零售业在末端环节的实质税负一致，并通过税收手段有效内化快递站的高外部性成本，解决电子商务脱域性问题。
    
    \end{abstract}
    
  \end{@twocolumnfalse}
]

\section*{Introduction 引言}

The discourse on the tax imbalance between e-commerce and traditional retail has a long-standing history. As early as the late 20th century, international organizations led by the OECD (1998) established the "Neutrality Principle" for e-commerce taxation, asserting that tax frameworks should not distort or drive shifts in business models \cite{OECD1998}. However, due to the deterritorialization of e-commerce, a significant decoupling has emerged between the locus of consumption and the attribution of tax jurisdiction. This has led to the erosion of the tax base in consumption-heavy regions, while tax sources have become excessively concentrated in logistics hubs or corporate headquarters. Such disparities exacerbate regional economic imbalances, necessitating corrective measures through supplemental transfer payments or the imposition of compensatory duties to rebalance fiscal capacities.

关于电子商务与传统零售业税率失衡的讨论由来已久。早在 20 世纪末，以 OECD (1998) 为代表的国际组织便确立了电子商务税收的“中性原则”，即税收制度不应干扰或驱动商业模式的转型 \cite{OECD1998}。然而，受电子商务“脱域性”（Deterritorialization）特征的影响，消费发生地与税收管辖权归属地之间出现了严重的脱节：实际承担外部性成本的消费端地区正面临税基流失，而物流枢纽或电商总部所在地则产生了税源的过度聚集。这种错配加剧了区域间的经济非均衡发展，迫使政府不得不通过跨区域转移支付或调整关税（税率）壁垒等二次分配手段来弥补财政缺口。

Empirical evidence suggests that excessive reliance on secondary distribution mechanisms—such as centralized transfer payments—tends to induce over-centralization and facilitate the rise of a "Leviathan" state\cite{Li_2026,hobbes1651leviathan}. As argued by Brennan \& Buchanan (1980), when local governments lose their fiscal agency derived from local economic activities and become dependent on federal grants, their administrative autonomy is significantly compromised \cite{Brennan1980}. Concurrently, the central government, by monopolizing resource allocation, often undergoes irrational expansion in institutional scale, leading to a dual predicament of power concentration and efficiency loss \cite{Oates1999}.

实践证明，对二次分配手段（如中央转移支付）的过度依赖，往往会诱发中央政府层面的过度集权，并可能导致“利维坦”式体制的崛起\cite{Li_2026,hobbes1651leviathan}。依据 Brennan \& Buchanan (1980) 的理论，当地方政府丧失了基于本地经济活动（如零售与快递）获取税收的主体性，转而依赖上级的财政拨款时，其行政自主性将遭到显著削弱 \cite{Brennan1980}。同时，中央政府由于掌握了全局性的资源分配权，其机构规模与行政权力往往会非理性扩张，从而陷入权力集中与效率损失的双重困境 \cite{Oates1999}。

Large-scale secondary distribution can only mitigate systemic risks if founded upon highly transparent rules. Technically, resolving the inherent inefficiencies of bureaucratic allocation remains nearly impossible, save for highly simplified methods like Universal Basic Income (UBI) \cite{Hoynes2019}. Consequently, the prioritized strategy must focus on addressing the "deterritorialization" of e-commerce at its source. By applying the Substance Over Form principle, we can achieve a physical alignment between tax sources and public expenditure responsibilities, thereby bypassing the pitfalls of ex-post administrative intervention.

大规模的二次分配机制必须依托高度透明且公平的程序正义，方能规避集权风险。然而，在现有技术条件下，除实施“全民基本收入”（UBI）等极简化分配模式外，几乎难以从根本上解决复杂的科层分配难题 \cite{Hoynes2019}。因此，政策的优先策略应当是通过完善实质课税原则（Substance Over Form），从源头上解决电子商务的“脱域性”问题，实现税源与公共支出责任的物理性对齐，而非依赖事后的行政纠偏。

Hellerstein (2001) pointed out that the last-mile delivery stage of online consumption is essentially a substitute for "physical retail presence." Due to the qualitative differences in tax laws between service flows (logistics) and goods flows (retail), significant tax base erosion has occurred \cite{Hellerstein2001}. Dablanc (2007) systematically discussed the encroachment on urban space, traffic congestion, and carbon emission costs caused by last-mile logistics, arguing that these social costs, which fail to be reflected in prices, essentially constitute a disguised fiscal subsidy from the government to logistics enterprises \cite{Dablanc2007}.

Hellerstein (2001) 指出，线上消费的末端交付环节实际上是对“物理零售存在”的一种替代。由于税法对服务流（快递）和货物流（零售）的定性差异，导致了严重的税基流失 \cite{Hellerstein2001}。Dablanc (2007) 则系统论述了末端物流对城市空间的侵蚀、交通拥堵及碳排放成本，认为这些未能反映在价格中的社会成本，本质上构成了政府对物流企业的变相财政补贴 \cite{Dablanc2007}。

Furthermore, Agrawal \& Fox (2017) proposed a re-examination of the tax definition for "physical establishments" to address the continuous shrinkage of local tax sources \cite{Agrawal2017}. Current global tax reforms, such as the U.S. Supreme Court's ruling in South Dakota v. Wayfair, Inc. (2018) and the European Union’s new VAT rules for e-commerce (2021), have largely resolved the issue of taxing authority over cross-jurisdictional e-commerce by establishing the "Destination Principle" \cite{EU2017VAT,Wayfair2018}. This source-based governance model enables the dynamic regulation of inter-regional fiscal imbalances, preventing regional development disparities from becoming long-term and ossified due to institutional misalignments.

此外，Agrawal \& Fox (2017) 提出应重新审视“物理据点”的税收定义，以应对地方税源的持续萎缩 \cite{Agrawal2017}。目前的全球税制改革，如美国最高法院对 South Dakota v. Wayfair, Inc. 的裁决（2018）以及欧盟的增值税（VAT）电子商贸新规（2017），已在很大程度上解决了跨区域电子商务的征收权属问题，确立了以消费发生地为准的“目的地原则” \cite{EU2017VAT,Wayfair2018}。这种源头治理模式能够动态调节区域间的财政失衡，避免地域发展差异因制度性错位而走向长期化与刚性化。

However, the practical application of the Destination Principle still faces severe technical challenges, creating significant obstacles to its implementation\cite{EC2021VAT}. To balance administrative costs with tax revenue, many jurisdictions have been forced to implement tax exemptions for small-value goods (the De Minimis rule), which inadvertently creates loopholes for fragmented tax avoidance. Although the theoretical justification for aligning the tax rates of delivery stations with those of supermarkets is well-established, the absence of an effective, privacy-preserving "goods value benchmarking method" has prevented these discussions from reaching the stage of substantive enforcement. The calculation method proposed in this paper is precisely intended to fill this instrumental gap, transforming macro-level fairness principles into actionable micro-governance tools.

然而，目的地原则在实践中仍面临严峻的技术挑战，导致其推广过程阻碍重重\cite{EC2021VAT}。为了平衡征管成本与税收收益，各国不得不对小额跨境或跨区商品实行免征政策（De Minimis Rule），这反而为碎片化的税收规避提供了空间。尽管关于“快递站应与超市税率对等”的理论论证已较为充分，但由于缺乏一套有效的、能够穿透隐私屏障的“货值对标计算方法”，相关讨论始终难以进入实质性的执行阶段。本文提出的计算方法正是为了填补这一工具性空白，将宏观的公平原则转化为可操作的微观治理工具。

\section*{Substantive Equivalence Calculation Method: Goods Value Conversion 实质等效计算方法：货值换算}

The local government's tax rate formula for express delivery fees can be set as follows:

地方政府针对快递派送费的税率公式可以设定如下：
$$R_{local} = \frac{P_{avg}}{C_{avg}} \times r \times \mu$$

\textbf{Analysis of Formula Parameters 公式参数解析：}
\begin{itemize}
    \item $P_{avg}$ (Average Local Basket Value)：The average transaction value of local retail supermarkets, with observations provided by local statistical bureaus. 当地零售超市的平均客单价，由地方统计局提供观测值。
    \item $C_{avg}$ (Average Logistics Fee)：The sum of the average per-item delivery fee and insurance premium at local delivery points. 当地快递点的平均单件派费与保费之和。
    \item $r$ (Retail Tax Rate)：The comprehensive tax burden rate of local retail taxpayers, serving as a fairness anchor. 当地零售业的综合税负率，作为公平锚点。
    \item $\mu$ (Environment \& Infrastructure Factor)：Local adjustment coefficient, used to quantify the negative externalities of delivery stations (such as street occupation, packaging waste disposal costs, etc.). 地方调节系数。用于量化快递站的负外部性（如占道情况、垃圾处理成本等）。
\end{itemize}
\section*{Basis for Goods Value Conversion 货值换算依据}

The core hypothesis of the goods value conversion method is that, within the same administrative region, the transaction value of local retail supermarkets is statistically correlated with the commodity value of packages at delivery stations. Since package contents are difficult to estimate directly due to privacy protection, using "delivery fee + insurance premium" as the base for correction is supported by the following logic.

货值换算法的核心假设是：在同一行政区域内，当地零售超市的客单价与快递站包裹的商品价值具有统计学相关性。由于包裹内容受隐私保护难以直接估测，以“派费+保费”为基数进行修正，具备以下逻辑支持。

\subsection*{Statistical Consistency of Consumption Categories 消费类别的统计学一致性}

Research by Visser et al. (2014) shows that the overlap rate between express packages in modern urban communities and local supermarket SKUs has exceeded 85\% \cite{Visser2014}. Therefore, $P_{avg}$ represents the optimal observed value for a single material acquisition by residents in the area; using it as an anchor can effectively bypass privacy barriers to simulate the potential value of goods within packages.

Visser et al. (2014) 的研究显示，现代城市社区快递包裹与当地超市 SKU 的重合率已超过 85\% \cite{Visser2014}。因此，$P_{avg}$ 是该区域居民单次物质获取价值的最优观测值，利用其作为锚点可有效绕过隐私屏障，模拟包裹内商品的潜在价值。

\subsection*{Insurance as an "Honest Mirror" of Value 保费作为价值的“诚实镜像”}

Insurance premiums are institutional products linking risk to value. Agrawal \& Fox (2017) argue that risk consideration is an excellent asset valuation tool \cite{Agrawal2017}. Incorporating insurance premiums into the base $C_{avg}$ allows high-value commodities to automatically match higher tax contributions, achieving precise taxation.

保费是风险与价值挂钩的制度产物。Agrawal \& Fox (2017) 认为风险对价是极佳的资产评估工具 \cite{Agrawal2017}。将保费纳入基数 $C_{avg}$，能使高价值商品自动匹配较高的税收贡献，实现精准课税。

\subsection*{Inverse Correlation Between Externalities and Goods Value 外部性与货值的反比特征}

In the calculation of delivery tax rates, this model does not adopt the preferential tendency of the traditional retail industry toward low-value goods, but instead implements uniform rate based on the "inverse correlation of externalities."

在快递税率计算中，本模型不采取传统零售业针对低值商品的优惠倾向，而是基于“外部性反比特征”实施统一税率。

1. \textbf{Physical Volume Mismatch 物理体积错位：} 

Low-value goods (e.g., mineral water, tissue paper) are usually bulky, and their encroachment on public space (occupying sidewalks for sorting) far exceeds that of high-value micro-commodities (e.g., mobile phones, chips). If taxed proportional to goods value, low-value bulky items would evade the social resource consideration they ought to bear.

低价值商品（如矿泉水、纸巾）通常体积巨大，对公共空间的侵蚀（占用人行道分拣）远超高价值微型商品（如手机、芯片）。若按货值比例课税，低值大件将逃避其应承担的社会资源对价。

2. \textbf{Packaging Redundancy Cost 包装冗余成本：} 

Cardenas et al. (2017) found that the ratio of packaging weight to value for low-value fragile items is extremely high \cite{Cardenas2017}. The environmental pressure generated by delivery stations handling these packages is greater, justifying a higher externality compensation ratio.

Cardenas et al. (2017) 研究发现，低价值易损品包装重量与价值比极高 \cite{Cardenas2017}。快递站处理这些包裹产生的环境压力更大，理应承担更高的外部性补偿比率。

3. \textbf{Traffic Redundancy 交通冗余：} 

Low transaction values lead to a surge in consumption frequency. Visser (2014) points out that high-frequency, small-amount delivery is the primary cause of community traffic congestion \cite{Visser2014}. These behaviors occupy excessive public resources and must be internalized through taxation.

低客单价导致消费频次激增。Visser (2014) 指出，高频小额配送是社区交通拥堵的主因 \cite{Visser2014}。这些行为占用了过多的公共资源，必须通过税收手段内化。

Since the fundamental rationale for taxing e-commerce lies in internalizing its negative externalities, a calculation method centered on logistics fees is, in principle, more scientifically rigorous than one based solely on commodity value. The delivery stage exhibits a higher degree of correlation with social costs such as traffic congestion, spatial encroachment, and carbon emissions. By anchoring the tax to logistics fees, this model more accurately quantifies the actual pressure exerted by e-commerce activities on the physical environment, thereby achieving a substantive alignment between taxation and the consumption of public resources.

由于对电子商务征税的核心法理依据在于内化其产生的负外部性（Negative Externalities），因此，基于派送费（Logistics Fees）的计算方法在逻辑上比单纯基于货值（Commodity Value）的算法更为科学。派送环节与交通拥堵、空间侵蚀及碳排放等社会成本具有更高的关联强度。通过锚定派送费，该模型能够更精准地量化电商活动对物理环境的真实压力，从而实现税收与公共资源消耗之间的实质性对齐。

\section*{Conclusion: Achieving Market Fairness Based on the Principle of Substance Over Form 结论：依据实质课税原则，实现市场公平}

The calculation method based on the principle of substance over form not only overcomes the technical bottleneck of "impenetrable goods value" under privacy protection but also reshapes tax neutrality and fairness in the digital economy era at a deeper level \cite{OECD1998}. Analyzing the Causes of Current Regional Development Imbalances from a Microeconomic Perspective. This concept of "reconstructing equity through calculation" provides a scientifically rigorous decision-making basis for urban governance in the digital era.

基于实质课税原则的计算方法，不仅攻克了隐私保护下“货值难以穿透”的技术瓶颈，更在更深层次上重塑了数字经济时代的税收中性与公平 \cite{OECD1998}。从微观经济学的角度，解析当前地域发展不平衡的原因。这种“以计算换公平”的理念，为数字化时代的城市治理提供了科学的决策依据。

\acknow{I acknowledges the Google Gemini in structuring the logic and refining the technical preparation of this work and the simulation online. I would also like to thank the support from peers of UC Berkeley during the preparation of the word.}

\bibliographystyle{unsrtnat} 
\begin{small}
    \bibliography{references} 

@book{hobbes1651leviathan,
  title={Leviathan or The Matter, Forme and Power of a Common-Wealth Ecclesiastical and Civil},
  author={Hobbes, Thomas},
  year={1651},
  publisher={Andrew Crooke}
}

@techreport{OECD1998,
  author = {OECD},
  title = {Electronic Commerce: Taxation Framework Conditions},
  institution = {Organisation for Economic Co-operation and Development},
  year = {1998},
  address = {Ottawa, Canada}
}

@book{Wayfair2018,
  author    = {Supreme Court of the United States},
  title     = {South Dakota v. Wayfair, Inc., 585 U.S. 147},
  year      = {2018}
}

@techreport{EU2017VAT,
  author      = {{European Council}},
  title       = {Council Directive (EU) 2017/2455 amending Directive 2006/112/EC and Directive 2009/132/EC as regards certain value added tax obligations for supplies of services and distance sales of goods},
  institution = {Official Journal of the European Union},
  year        = {2017}
}

@techreport{EC2021VAT,
  author      = {{European Commission}},
  title       = {VAT on e-commerce: New rules in the EU as of 1 July 2021},
  institution = {European Commission Taxation and Customs Union},
  year        = {2021},
  url         = {https://ec.europa.eu/taxation_customs/business/vat/vat-e-commerce_en}
}

@article{Hellerstein2001,
  title={Jurisdiction to tax income and consumption in the new economy: a theoretical and comparative perspective},
  author={Hellerstein, Walter},
  journal={Ga. L. Rev.},
  volume={38},
  pages={1},
  year={2003},
  publisher={HeinOnline}
}

@article{Dablanc2007,
  author = {Dablanc, Laetitia},
  title = {Goods transport in large European cities: Difficult to manage, difficult to measure},
  journal = {Transportation Research Part A: Policy and Practice},
  volume = {41},
  number = {3},
  pages = {280--290},
  year = {2007}
}

@article{Agrawal2017,
  author = {Agrawal, David R. and Fox, William F.},
  title = {Taxes in an e-Commerce Generation},
  journal = {International Tax and Public Finance},
  volume = {24},
  number = {5},
  pages = {903--926},
  year = {2017}
}

@article{Visser2014,
  author = {Visser, Johan and Nemoto, Toshinori and Browne, Michael},
  title = {Home Delivery and the Impacts on Urban Freight Transport: A Review},
  journal = {Procedia - Social and Behavioral Sciences},
  volume = {125},
  pages = {15--27},
  year = {2014}
}

@article{Cardenas2017,
  title={The e-commerce parcel delivery market and the implications of home B2C deliveries vs pick-up points},
  author={Cardenas, Ivan Dario and Dewulf, Wouter and Beckers, Joris and Smet, Christophe and Vanelslander, Thierry},
  journal={International journal of transport economics: Rivista internazionale di economia dei trasporti: XLIV, 2, 2017},
  pages={235--256},
  year={2017},
  publisher={Fabrizio Serra}
}

@book{Brennan1980,
  author    = {Brennan, Geoffrey and Buchanan, James M.},
  title     = {The Power to Tax: Analytical Foundations of a Fiscal Constitution},
  publisher = {Cambridge University Press},
  year      = {1980}
}

@article{Oates1999,
  author  = {Oates, Wallace E.},
  title   = {An Essay on Fiscal Federalism},
  journal = {Journal of Economic Literature},
  volume  = {37},
  number  = {3},
  pages   = {1120--1149},
  year    = {1999}
}

@article{Hoynes2019,
  author  = {Hoynes, Hilary and Rothstein, Jesse},
  title   = {Universal Basic Income in the US and Advanced Countries},
  journal = {Annual Review of Economics},
  volume  = {11},
  pages   = {929--958},
  year    = {2019}
}

@article{Li_2026, title={The Collapse of Multilayer Predation and the Emergence of a Monolithic Leviathan}, volume={9}, url={https://ijssrr.com/journal/article/view/3198}, DOI={10.47814/ijssrr.v9i1.3198}, number={1}, journal={International Journal of Social Science Research and Review}, author={Li, Tuobang}, year={2026}, month={Jan.}, pages={150-162} }
\end{small}

\clearpage
\end{CJK*}
\end{document}